# Hidden time interpretation of quantum mechanics and "no - protocol" argument

P. V. Kurakin[*]

[*]Moscow Institute of Physics and Technology

## ABSTRACT

Previously suggested hidden time interpretation of quantum mechanics allows to reproduce the same predictions as standard quantum mechanics provides, since it is based on Feynman many – paths formulation of QM. While new experimental consequences of this interpretation are under investigation, some advantages can be enumerated. (1) The interpretation is much field theoretic – like in classical sense, so it is local in mathematical sense, though quantum (physical) non-locality is preserved. (2) The interpretation is based on one type of mathematical objects, rather than two different (Hilbert space vectors and operators). (3) The interpretation, as it was argued, overcomes the problem of hidden variables in a radically new way, with no conflict to Bell's theorem. Recently an important argument against hidden variables – like formulations of quantum theory was risen – "no protocol" argument. It is argued in the paper, that hidden time interpretation successfully overcomes this argument.

**Keywords:** quantum mechanics, quantum entanglement, hidden time, no – protocol argument

## 1. INTRODUCTION

### 1.1 The core algorithm of hidden time interpretation of quantum mechanics

The core of hidden time interpretation of quantum mechanics [1 - 4] can be shortly formulated as follows:

1. A quantum particle travels from the source emitter to destination detector (all detectors) by all classically possible paths, exactly as in Feynman's formulation of quantum mechanics. This happens in what is called hidden time. For the start understanding it is enough to accept that the latter is not the same as physical (observed) time. That is, don't try to imagine how particle *physically* travels all paths in physical time.

2. Such pathways are not just ordinary trajectories; they "stay" in space (in hidden time!) as oriented "roadways". I refer to these paths as *scout signals* (the particle searches for all possible detectors). The notion of signal is very specific here, because it is both the signal and its container which matters. Such signals are not unusual for those familiar with "artificial life" models by Christopher Langton [5].

3. While traveling along each path, a scout signal calculates the quantum mechanical amplitude for this path. For a photon it is equal to $\exp(i\varphi)$, $\varphi = L/c$, where $L$ is optical length of the path.

4. When ("when" in hidden – time sense) all possible scout signals have reached a certain destination detector, all amplitudes are summed. So, the full quantum mechanical amplitude of reaching any possible detector is found. This is so in the original version of hidden time interpretation. See below some new nowadays speculations about this point.

5. Next (all in hidden time) a new kind of signals travel backward from each detector to the source of the particle. These are *query signals*. They all have constant amplitude value, the one calculated at the loci of detectors as described in previous step.

6. Query signals come to the source. For each possible detector, they travel by *all* paths, which were constructed by scout signals at step 1. Again, this must not be unusual for those familiar with cellular automata – based models of living cells. In the original version of the model query signals corresponding to each single detector, compete to each other while traveling to the source. Finally a single copy of a query signal enters **per detector** enters the source.

7. There happens a kind of lottery at the locus of the source of the considered particle: query signals compete, for the only one to win. The probability to win the lottery is proportional to $|\psi|^2$, where $\psi$ is the amplitude of corresponding detector. The winner signal travels again to corresponding detector, while others receive "rejection signals".

These points constitute the mathematical core of the hidden time interpretation but not the physical, which is discussed below. The core assertion for mathematical part is: hidden time interpretation is equivalent to quantum mechanics in predictions as soon as it is equivalent Feynman's path integrals formulation of QM.

**1.2 Relation of hidden time and physical time and sewing procedure**

The algorithm above, in fact, implements feedback from *future* to present. Such idea is not new: backward causation is broadly discussed nowadays [6]. Hidden time interpretation also adds "backward causation" to Feynman path integrals. But the radical innovation hidden time interpretation provides is a hypothesis on the physical nature of time. The interpretation distinguishes "hidden" time and "physical" (or, observed) time. All kind of signals described above propagate in hidden time. Physical time "ticks" at the *moment* (in hidden time) when the winner detector receives the confirmation signal only. This can be understood as analogy to Poincare sections of a quasi – periodical trajectory. I refer to this map of hidden time instants to physical time instants as *sewing procedure*.

Such map makes sense only if we clearly define what physical (observed) time is. The meaning of physical time in connection with quantum mechanics was discussed, but no common understanding is reached up to date [7 - 9]. It is clear that physical time is something that can be *observed* (in other terms, detected or measured), but in quantum mechanics physical time is just a parameter in equations, not an observable. The paper [1] argues that observed physical can be defined as a (properly normalized) count of detected elementary events, say photon absorptions; it also argued that defined so physical time is fully consistent with suggested sewing procedure.

In [2] it is also argued that though hidden time is obviously an *absolute* kind of "time", the detected events, which constitute the fabric of physical space-time, satisfy Lorenz transformations. I.e. absolute hidden time provides Minkowski space time *for detected events*.

## 2. NO – PROTOCOL ARGUMENT

**2.1 "No – protocol" theorem and its assumptions**

As it was argued at the very claiming of hidden time interpretation of quantum mechanics, the latter *does not* conflict Bell's theorem, though scout, query, confirmation and rejection signals are in fact hidden variables. The reason is that Bell's theorem explicitly uses assumption that statistical properties of hidden variables *can not depend* on macroscopical setup parameters of remote detectors involved in EPR – like experiments. Hidden time interpretation involves variables (describing all kinds of signals), which *do* obtain values dependent on detectors' parameters, since the scout signals reach detectors before (in hidden time) final choice is made and "read" the setup parameters of these detectors (which stay "frozen" in hidden time).

So, the problem of Bell's theorem is that it explicitly assumes some *very specific* type mathematically possible hidden variables whereas the theorem is interpreted *too wide*, as if it assumes any type of hidden variables.

Another very interesting counter – argument against hidden variables and "back in time causation" was recently raised [10]. It worth citing that paper vast:

"…Two explanations appear most frequently in literature for the queer interdependence between the behavior of space-separated particles: the local hidden variables (LHV) assumption and the non-local hidden variables (NLHV) based on the idea of faster-than-light or backward-in-time messages – the so-called "signals". It is supposed that the "signals" are exchanged among particles, and carry information so that the particles can come to an "agreement" on what results to produce in measurements.

The LHV assumption was for long proved unable to explain the experimental results. On the other hand, the "signals" contradict the theory of relativity, according to which no objects can move faster than light.

…If one assumes that the experimental results are obtained on base of message-passing, one implicitly assumes that there exists a *protocol of communication*.

…Five *basic assumptions* are adopted:

*a*) The apparatus is ideal – no losses.

*b*) Outcomes of measurements are consequences of the properties of the quantum objects, not of the detectors.

*c*) Experimental results are generated during the measurement – at the interaction with the classical detector – not before. Therefore a message containing information about a measurement is issued during that measurement.

*d)* A message comprises only local information.

*e)* Messages are broadcast; from the region of the measurement of a given particle, the message is emitted to all the space and covers all the other regions…"

Next, it is shown in the paper that constructing a communication protocol for entangled particles meets unavoidable difficulties and controversies unresolved in principle.

**2.2 Hidden time escapes assumptions of "no - protocol" theorem**

As it can be easily seen, this argument suffers the same problem as Bell's theorem. A *very specific* type of "signaling" is assumed and we can't generalize the "no - protocol" theorem to all *mathematically possible* types of signals. Hidden time signals violate assumptions (c), (d) and (e):

- Experimental results are generated *before* observed measurement (in hidden time) at the loci of the source(s), but *no when,* from physical time viewpoint.

- Hidden time signals carry the information, which is local only formally: these amplitudes at loci of corresponding detectors, but the very values of amplitudes are obtained as a result of length calculation along multiple paths. That is, this information can be assumed both local and non-local.

- Messages that hidden time signals carry are *not* broadcasted: they propagate along paths to selected locations.

## 2.3 A comment on basic assumptions of "no - protocol" assumption

It looks very much, though it is not explicitly asserted in [10], as if the theorem *implicitly* suggests that quantum mechanics is *fundamental* theory of physics. Assumptions above look reasonable if we admit the same. I see the problem with quantum mechanics not in reducing quantum mechanics to another well-known but non-quantum physics, i.e. to *avoid* quantum mechanics at all.

The problem *can be* understood as to *obtain* quantum mechanics (and possibly some violations of the latter and\or new predictions) from a *more* fundamental model. Hidden variables, as originally suggested by A. Einstein, are *beyond* quantum mechanics. They do not have sense out of this concept.

This is why I argue that when considering the idea of hidden variables, we should concentrate on *mathematical* aspects. Hidden variables must not behave *physically*. Physical dynamics *can be* what is observed only, as a kind of foam over hidden variables dynamics. In this paradigm only the resulting observed dynamics emerged from a more fundamental dynamics must be physical.

## 3. MODIFIED HIDDEN – TIME ALGORITHM

### 3.1 The problem

Still the model constituting hidden time interpretation should be modified. The problem of initial version is the following (pointed by Sofia Wechsler in private communication). The hidden time algorithm suggested in [3] takes for granted that no quantum entanglement emerges without physical interaction actually present between entangling particles and some devices involved in the experiment. Say, in Hong - Ou – Mandel experiment photons must follow the same beam splitter to become entangled. Still, as it is shown in [11], quantum mechanics allows entanglement to emerge for systems where there's no actual physical interaction for involved particles, but such an interaction is possible in the experiment.

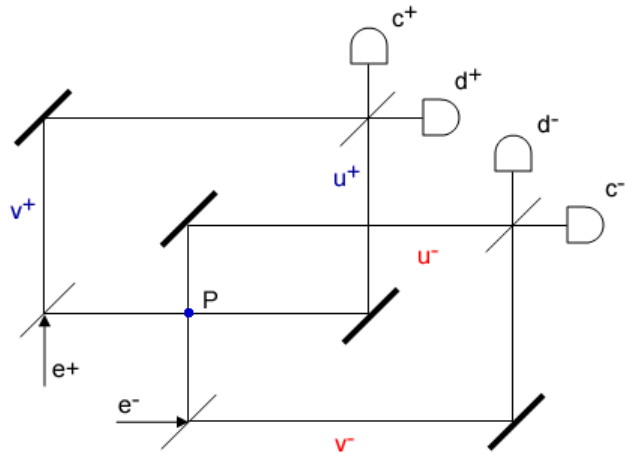

Fig. 1. Hardy's gedankenexperiment [11]. Electron and positron can annihilate at point P which generates a gamma - quantum.

In Hardy's gedankenexperiment an electron and a positron enter Mach – Zehnder interferometers which have common point P on the intersection of particles' pathways (Fig. 1). The initial electron and positron states transfer as $|s^{\pm}\rangle \to \left(1/\sqrt{2}\right)\left(i|u^{\pm}\rangle+|v^{\pm}\rangle\right)$ at 1st beamsplitters of interferometers. At the 2nd beamsplitters states split analogously. If electron and positron meet at the point P, they annihilate: $|u_+\rangle|u_-\rangle \to |\gamma\rangle$. So, when we finally sum the states that emerge we get the state

$$|s^+\rangle|s^-\rangle \to \frac{1}{4}\left(-2|\gamma\rangle - 3|c^+\rangle|c^-\rangle + i|c^+\rangle|d^-\rangle + i|d^+\rangle|c^-\rangle - |d^+\rangle|d^-\rangle\right).$$

The question is how hidden time interpretation can provide an algorithm to reproduce probability amplitudes the same as this state does. As it was pointed above, the algorithm in [3] produces entangled states but it is bases on simultaneous interaction of entangling photons with a beamsplitter (as in the Hong - Ou – Mandel experiment); while Hardy's experiment gives entanglement without mandatory interaction to happen.

### 3.2 The Solution (preamble)

1st, let us come back to many – paths formulation of QM by Feynman to calculate two – partite amplitudes. Say, experimental output $|c^+\rangle|c^-\rangle$ can be classically achieved by 3 two – partite pathways (Fig 2(a), (b), (c)).

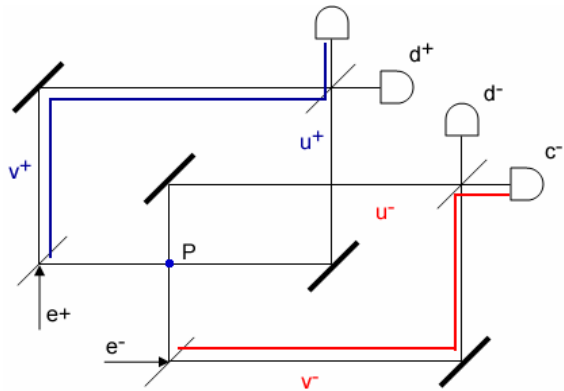

Fig. 2(a) Single – particle paths that lead to $c^+c^-$ output

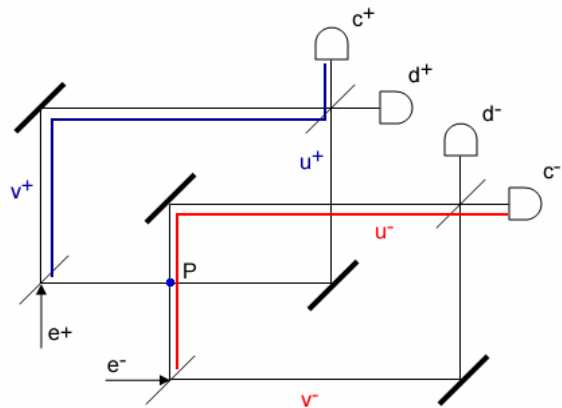

Fig. 2(b) Single – particle paths that lead to $c^+c^-$ output

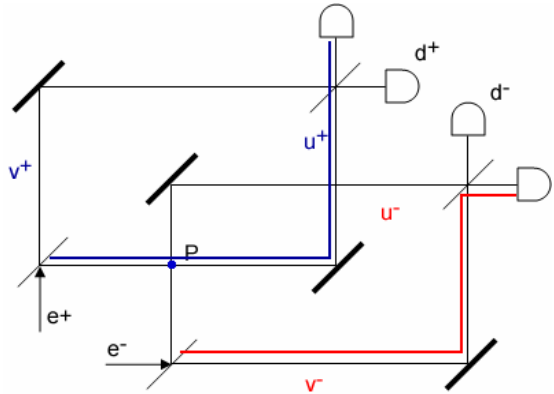

Fig. 2(c) Single – particle paths that lead to $c^+c^-$ output

Summing path integrals for all three two – partite paths we get amplitude $A_{C^+C^-} = -\dfrac{3}{4}$.

### 3.3 The Solution (the algorithm)

- As before (in initial version of hidden – time algorithm), we put correspondence between each single – partite path and a scout signal: each particle forms a set of its own (1 – particle) paths to its possible outcomes (detectors). This allows the whole multiparticle system to discover where common interaction point is (Fig. 3).

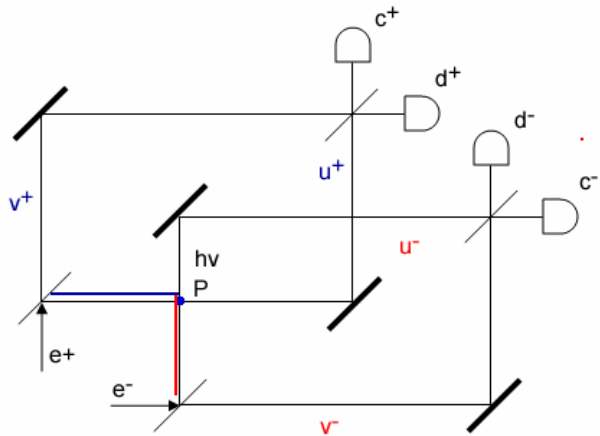

Fig. 3. Some of single – partite paths discover interaction locus, when meet.

- Each such path is bent and modified so as to enter common interaction point (see comment below for more than 1 common interaction location) – see Fig.4

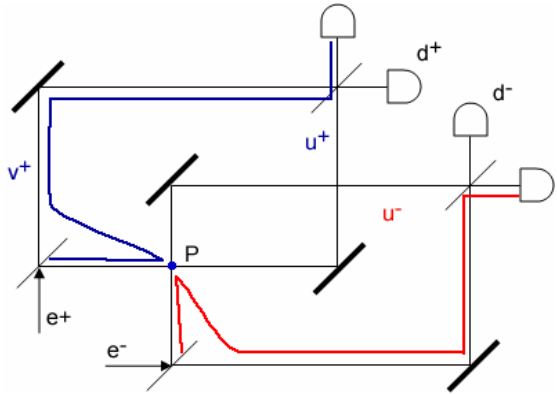

Fig. 4. Single – particle paths "bend" so as to enter common interaction point

- All 2 – particle paths are formed by tensor (Descartes) multiplication and correspondent 2 – particle amplitudes are calculated (Fig. 5). In hidden time, "to calculate amplitude" means that a signal (a wave) circulating *along* the path acquires that amplitude.

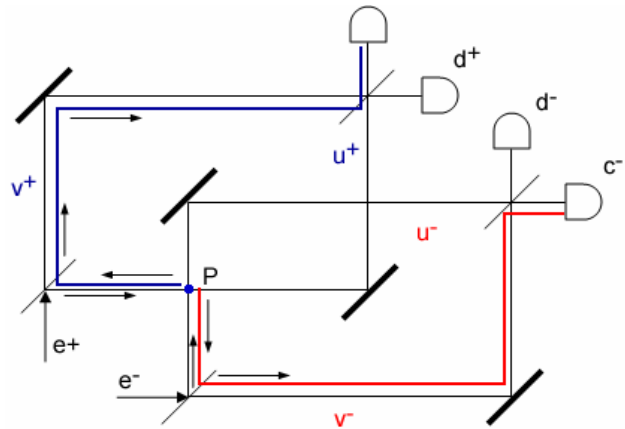

Fig. 5. Two – partite paths are formed of single-particle paths so that they include parts

- Amplitudes for indistinguishable outcomes are added, as is standard for QM. Identity of paths leading to the same results is yet to be discussed.
- Finally, all **2-particle** paths (= amplitudes) "meet" <u>*at the common interaction point*</u> and "compete": a lottery selects 1 (2-particle!) outcome, as in the initial version of the hidden time algorithm described in [3]. Each amplitude's chance of to win is squared amplitude, with full agreement with standard QM.

## CONCLUSION

So I suggest that "no - protocol" argument is easily met by hidden – time paradigm. The key point, as it is pointed above, is to separate physical considerations from mathematical, and not to mix them. There are no any *physical* "signals" that coordinate observed quantum correlated properties of entangled particles; but this is not equal to a concept that quantum mechanics can not be *derived* from dynamics of underlying and non – observed (sub - physical) signal-based system.

Being non – observable is, of course, an additional hypothesis on the nature of underlying dynamical system we talk about (besides using signals), but is *fully* in agreement with current theory, since only many – partite events are physically observed, in principle. So extending the theory by introducing non – observed single particle events can not be treated as artificial.

I stand that such introducing of hidden time is a very *natural* way of thinking and a very natural continuation of development of QM.